\documentclass[twocolumn,prl,superscriptaddress]{revtex4}
\usepackage{amsmath,amssymb}
\usepackage{graphicx}
\usepackage{float}
\usepackage{subfigure}
\usepackage{verbatim}

\usepackage{amsfonts}
\usepackage{dcolumn}
\usepackage{bm}
\usepackage{color}
\usepackage[colorlinks,citecolor=blue]{hyperref}

\begin{document}
\hyphenpenalty=5000
\tolerance=1000

\title{
{Experimental demonstration of the Einstein-Podolsky-Rosen steering game based on the All-Versus-Nothing proof}
}

\author{Kai Sun}
\affiliation{Key Laboratory of Quantum Information, University of Science and Technology of China, CAS, Hefei, 230026, People's Republic of China}
\affiliation{Synergetic Innovation Center of Quantum Information and Quantum Physics, University of Science and Technology of China, Hefei, Anhui 230026, P. R. China}

\author{Jin-Shi Xu}
\email{jsxu@ustc.edu.cn}
\affiliation{Key Laboratory of Quantum Information, University of Science and Technology of China, CAS, Hefei, 230026, People's Republic of China}
\affiliation{Synergetic Innovation Center of Quantum Information and Quantum Physics, University of Science and Technology of China, Hefei, Anhui 230026, P. R. China}

\author{Xiang-Jun Ye}
\affiliation{Theoretical Physics Division, Chern Institute of Mathematics, Nankai University, Tianjin, 30071, People's Republic of China}

\author{Yu-Chun Wu}
\affiliation{Key Laboratory of Quantum Information, University of Science and Technology of China, CAS, Hefei, 230026, People's Republic of China}
\affiliation{Synergetic Innovation Center of Quantum Information and Quantum Physics, University of Science and Technology of China, Hefei, Anhui 230026, P. R. China}

\author{Jing-Ling Chen}
\email{chenjl@nankai.edu.cn}
\affiliation{Theoretical Physics Division, Chern Institute of Mathematics, Nankai University, Tianjin, 30071, People's Republic of China}
\affiliation{Centre for Quantum Technologies, National University of Singapore, 3 Science Drive 2, Singapore, 117543}

\author{Chuan-Feng~Li}
\email{cfli@ustc.edu.cn}
\affiliation{Key Laboratory of Quantum Information, University of Science and Technology of China, CAS, Hefei, 230026, People's Republic of China}
\affiliation{Synergetic Innovation Center of Quantum Information and Quantum Physics, University of Science and Technology of China, Hefei, Anhui 230026, P. R. China}

\author{Guang-Can Guo}
\affiliation{Key Laboratory of Quantum Information, University of Science and Technology of China, CAS, Hefei, 230026, People's Republic of China}
\affiliation{Synergetic Innovation Center of Quantum Information and Quantum Physics, University of Science and Technology of China, Hefei, Anhui 230026, P. R. China}

\begin{abstract}
Einstein-Podolsky-Rosen (EPR) steering, a generalization of the original concept of ``steering" proposed by Schr\"{o}dinger, describes the ability of one system to nonlocally affect another system's states through local measurements. Some experimental efforts to test EPR steering in terms of inequalities have been made, which usually require many measurement settings. Analogy to the ``All-Versus-Nothing" (AVN) proof of Bell's theorem without inequalities, testing steerability without inequalities would be more strong and require less resource. Moreover, the practical meaning of steering implies that it should also be possible to store the state information on the side to be steered, a result that has not yet been experimentally demonstrated. Using a recent AVN criterion for two qubit entangled states, we experimentally implement a practical steering game using quantum memory. Further more, we develop a theoretical method to deal with the noise and finite measurement statistics within the AVN framework and apply it to analyze the experimental data. Our results clearly show the facilitation of the AVN criterion for testing steerability and provide a particularly strong perspective for understanding EPR steering.
\end{abstract}


\maketitle

In 1935, Einstein, Podolsky and Rosen published their famous paper proposing a now well-known paradox (the EPR paradox) that cast doubt on the completeness of quantum mechanics \cite{epr1935}. To investigate the EPR paradox, Schr\"{o}dinger introduced the concept of ``steer'' \cite{schr1935}, now known as the EPR steering \cite{wjd2007}. As an asymmetric concept, EPR steering describes the ability of a system to nonlocally affect the states of another system through local measurements. EPR steering exists between the concepts of entanglement and Bell nonlocality \cite{bell1964}; these steerable states are a subset of the entangled states and a superset of Bell nonlocal states \cite{wjd2007}. A quantitative criterion for realizing EPR steering based on the uncertainty relation has been proposed~\cite{reid1989} and experimentally demonstrated~\cite{ou1992,bowen2003}, and the steerability of quantum states has been further formulated and characterized by general EPR steering inequalities \cite{cjwr2009}. This method, which usually requires many measurement settings, has been used to demonstrate the steerability of a class of Bell-local states, where states are still steerable even if they do not violate the Bell inequalities~\cite{saunders2010}. Recently, a new family of EPR steering inequalities based on entropic uncertainty relations have also been proposed and demonstrated experimentally~\cite{schneelochpra2013,schneelochprl2013}.

When characterizing Bell nonlocality, the strongest conflict between the predictions of quantum mechanics and the local-hidden-variable theory appears in the so-call All-Versus-Nothing (AVN) demonstration \cite{mermin1990}, in which the outcomes predicted by quantum mechanics occur with a probability of 0 and with a probability of 1 for the local-hidden-variable theory, and vice versa. In the AVN demonstration, inequalities are not needed \cite{adan200186,adan200187}, and so has been used to test nonlocality using a hyperentangled source \cite{yang2005,cinelli2005,vallone2007}. An AVN proof for EPR steering was recently proposed for two-qubit entangled states~\cite{chenjl2013}, in which the different pure normalized conditional states (NCS) in one qubit were used as a criterion along with a given projective measurement on the other. According to quantum mechanics, two different pure NCS should be obtained, while the local hidden state (LHS) model predicts that one cannot obtain two different pure NCS when the other qubit is performed by a projective measurement~\cite{chenjl2013}.

There is practical meaning in the concept of steering, which implies that it should be possible to store the state information on the side to be steered. Physically, Bob measures his qubit after receiving the measurement results sent by Alice. However, there has been no related experimental demonstration of this result. Therefore, in this work we propose a practical steering game using quantum memory and experimentally demonstrate it by employing the AVN criterion~\cite{chenjl2013}. The particle to be steered is initially stored in quantum memory. After measurement of the other particle, we can then check the states of the particle in the quantum memory and verify the steerability of the states they shared.

\begin{figure}
\centering
\includegraphics[width=0.45\textwidth]{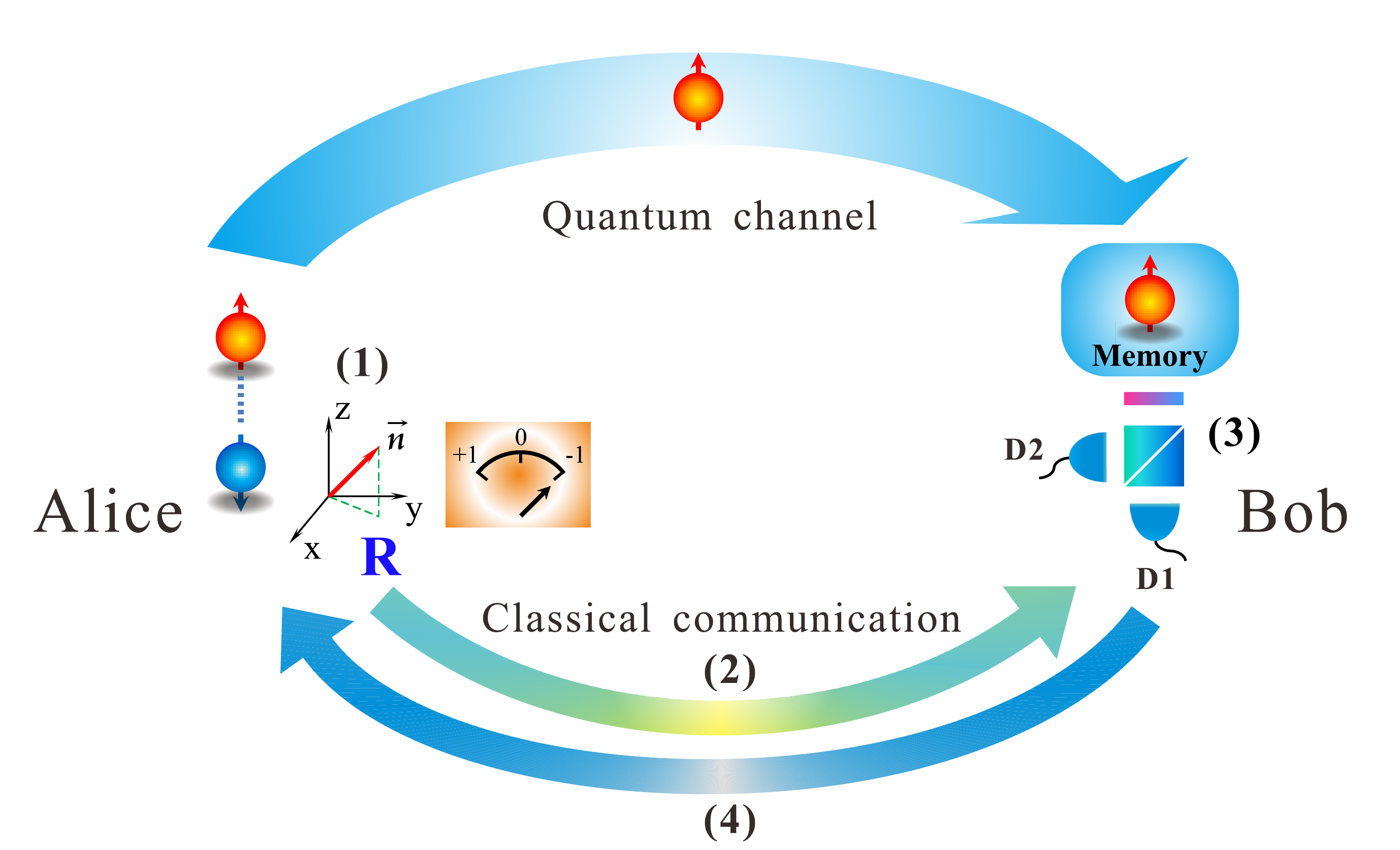}
\caption{{\bf Illustration of the EPR steering game.} {\bf (1)} Alice measures her own qubit along the direction $\vec{n}$ and obtains the outcomes of $+1$ and $-1$. {\bf (2)} Through a classical channel, Alice tells Bob her measurement outcome and the corresponding output state that Bob should obtain. {\bf (3)} Bob verifies the normalized conditional states. {\bf (4)} Alice and Bob implement a joint measurement. They determine the value of $\langle\mathcal{W}\rangle_{max}$ and compare it with the upper bound predicted by the LHS model ($C_{LHS}$).
}\label{theory}
\end{figure}

The EPR steering game is shown in Fig. ~\ref{theory}. Two-qubit entangled states are first prepared by one participant, Alice, who claims that these states are steerable. However, the other participant, Bob, does not trust Alice. Alice then starts the game by sending the steerable qubit to Bob through quantum channels, who stores it in quantum memory. To verify steerability, Alice then chooses a measurement direction $\vec{n}$ to make sure two different pure NCS are collapsed to the particle owned by Bob (denoted as $\rho_{B}^{\vec{n}}$). According to the measurement outcome, Alice tells Bob though classical communication which pure states ($|\phi\rangle_B$) he should obtain. Bob then reads the qubit from quantum memory, which is an indispensable part because the qubit through the quantum channel comes earlier than the signal from Alice via the classical channel, and checks it by projecting the qubit into the corresponding states and replies to Alice. The detector $D1$ is used to detect $|\phi\rangle_B$ with the probabilities denoted by $P_{+}$ and $P_{-}$ according to the outcomes $+1$ and $-1$ of Alice, respectively. The detector $D2$ is used to detect the states that are orthogonal to $|\phi\rangle_B$, where the corresponding probabilities are denoted by $P_{+}'$ and $P_{-}'$ which correspond to $\mathcal{W}_1$ and $\mathcal{W}_2$ in Ref.\,\cite{chenjl2013}. If two different pure NCS are obtained by Bob, i.e., the values of $P_{+}$ and $P_{-}$ are both equal to 1, and $P_{+}'$ and $P_{-}'$ are both equal to 0, the entangled states they shared are steerable. In general, there is no reason for Bob to agree with Alice that the initial state they shared is entangled. For example, Alice may cheat Bob, or there could be some sort of environmental disturbance that changes the state properties. To verify the result and rule out the possibility of cheating, Alice and Bob implement a joint measurement. It has been shown that a value of the equation \begin{equation}\label{delta}
\Delta=\langle\mathcal{W}\rangle_{max}- C_{LHS}
\end{equation}
 should further be checked to verify the steerability of the shared states even if two different pure NCS are obtained by Bob~\cite{chenjl2013}. In the equation, $\langle\mathcal{W}\rangle_{max}$ represents the maximal mean value of the joint operator $\mathcal{W}=|n^\bot\rangle\langle{n^\bot|}\otimes|\hat{n}_B\rangle\langle{\hat{n}_B|}$, with Alice measuring along the $n^\bot$ direction (perpendicular to $n$) and Bob measuring along $|\hat{n}_B\rangle=\cos\frac{\theta_B}{2} |0\rangle+\sin\frac{\theta_B}{2}e^{i\phi_B} |1\rangle$. Furthermore, the equation
 \begin{equation}\label{clh}
  C_{LHS}=\max_{n_B}{Tr(\rho_{AB}\cdot I\otimes|\hat{n}_B\rangle\langle{\hat{n}_B}|)}
  \end{equation}
  represents the upper bound predicted by the LHS model, where $I=\frac{1}{2}(|+n\rangle\langle +n|+|-n\rangle\langle -n|)$ is the identity operation of $\vec{n}\cdot \vec{\sigma}$ with $|\pm n\rangle$ being the eigenstates of $\vec{n}\cdot \vec{\sigma}$ and $\vec{\sigma}=(\sigma_x,\sigma_y,\sigma_z)$ representing the vectors of the Pauli matrices. If $\Delta>0$, the steering game is verified to be successful; while $\Delta\leq0$ indicates the steering game failed.

However, in practice, the measured states on Bob's side can never be sure to be indeed pure due to the effect of noise and finite measurement statistics. We develop a theoretical method to deal with the experimental errors within the AVN framework \cite{si}. Assuming the values of $P_{+}'$ and $P_{-}'$ whose results should be both $0$ in theory are $\epsilon_ 1$ and $\epsilon_ 2$, i.e.,
\begin{equation}\label{exp}
\begin{split}
P_+'=\epsilon_ 1,\,\ P_+=1-\epsilon_ 1, \\
P_-'=\epsilon_ 2,\,\ P_-=1-\epsilon_ 2
\end{split}
\end{equation} we prove that the shared state is steerable in the case of two settings if the following inequation is violated,
\begin{equation}\label{exp1}
 \Delta '=(OB-OG)_{min}\leq 0
\end{equation} where $OB$ is the length of a Bloch vector predicted by the LHS model, and $OG$ is the corresponding length determined by the experimental results of $P_+'$, $P_-'$, $\mathcal{W}_{max}$ with its probability $P_D$ and $\mathcal{W}'$ which represents the other eigenvalue of the $n^\bot$ direction relative to $\mathcal{W}_{max}$. The inequation (\ref{exp1}) is derived and discussed in detail in the supplementary material \cite{si}. Here we give a short discussion on the main idea of the criterion. According to the definition of steering, if a state is not steerable, then there is a LHS model to describe the conditional states on Bob's side after Alice's measurement. In the case of two measurement settings, it has been proved that four hidden states are enough to simulate the four conditional states on Bob's side \cite{AVNEx}. We show that these states can be mapped to states in the X-Z plane of a Bloch sphere. We further show that if there is not a LHS model on an isosceles trapezoid to represent the symmetrical conditional states, then there is not LHS model for the four conditional states on Bob's side. As a result, according to the geometry relationship between the states in the X-Z plane (corresponding figures can be found in the supplementary material), we can derive the criterion, i.e., if $OB>OG$, then the state we discussed is steerable. There are experimental errors in measuring the corresponding experimental values. We need to find the minimum value of $OB-OG$ by scanning the region given by the measured value with the corresponding errors. The final criterion is then given by the form of inequation (\ref{exp1}).

\begin{figure}
\centering
\includegraphics[width=0.45\textwidth]{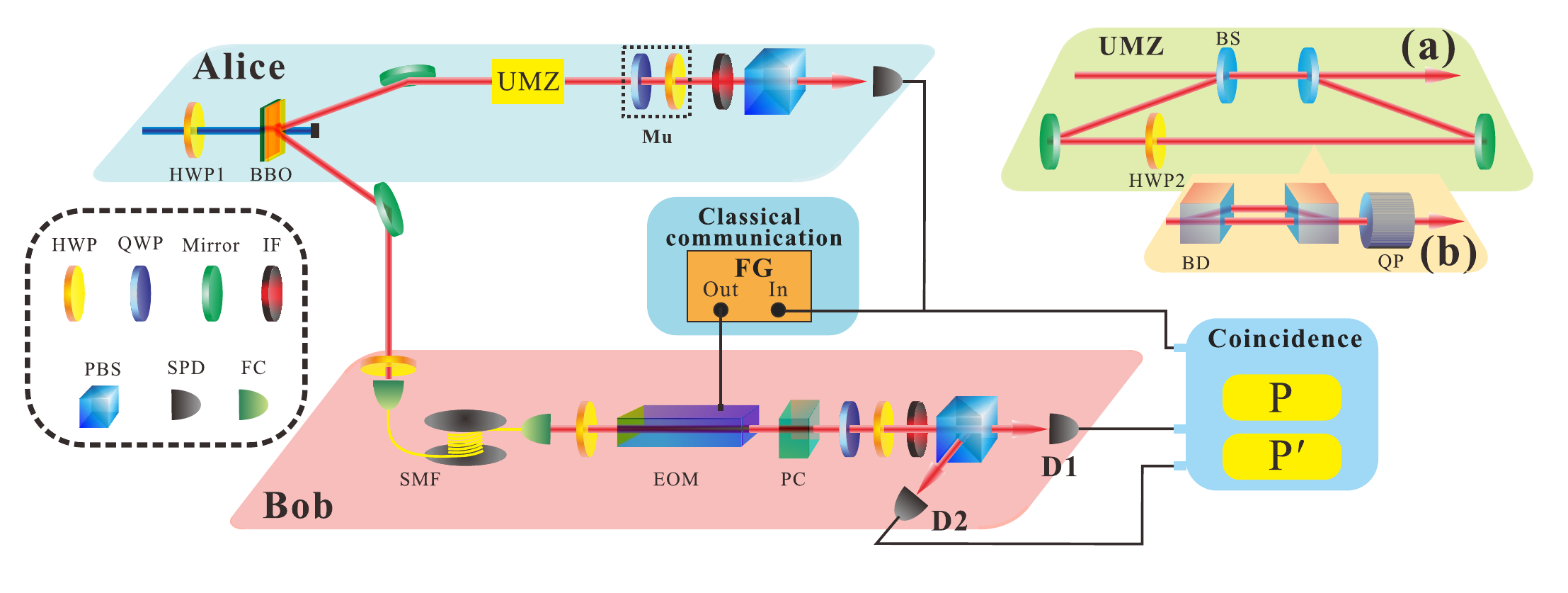}
\caption{{\bf Experimental setup.} The entangled photon pairs are produced via SPDC. An unbalanced Mach-Zehnder (UMZ) interferometer {\bf (a)} is employed to prepare the states $\rho_{1}$. The state $\rho_{2}$ is prepared by inserting UMZ {\bf (b)} consisting of two beam displacers (BD) and a quartz plate (QP) in the long arm of UMZ {\bf (a)}.
}
\label{setup}
\end{figure}

In our experiment, we prepare two kinds of polarization entangled states to demonstrate the EPR steering game
\begin{equation}\label{state}
\begin{split}
&\rho_1=\eta|\Psi(\theta)\rangle\langle\Psi(\theta)|+(1-\eta)|\Phi(\theta)\rangle\langle\Phi(\theta)|, \\
&\rho_2=\eta|\Psi(\theta)\rangle\langle\Psi(\theta)|+\frac{1-\eta}{2}(|HH\rangle\langle HH|+|VV\rangle\langle VV|),
\end{split}
\end{equation}
where $0\leq{\eta}\leq1$. Here, $|\Psi(\theta)\rangle=\cos{\theta}|HH\rangle+\sin{\theta}|VV\rangle$ and $|\Phi(\theta)\rangle=\cos{\theta}|VH\rangle+\sin{\theta}|HV\rangle$, where $|H\rangle$ denotes the horizontal polarization of the photons and $|V\rangle$ denotes the vertical polarization. Our experimental setup is shown in Fig. ~\ref{setup}. The entangled photon pairs are generated via spontaneous parametric down conversion (SPDC) \cite{kwiat1999}. The two-photon states $\rho_1$ and $\rho_2$ are prepared by Alice using the unbalanced Mach-Zehnder (UMZ) interferometer setup \cite{jsx2013} which is explained in detail in the supplementary material \cite{si}. The unit consisting of a quarter-wave plate (QWP) and a HWP, denoted as Mu, is used to set the measurement direction $\vec{n}$. Single-photon detectors (SPD) equipped with 3 nm interference filters (IF) are used to count the photons. The electric signal from the SPD on Alice's side is divided into two parts. One part is used as the trigger signal for the function generator (FG) while the other part is sent to the coincidence unit. The photon sent to Bob is then delayed by a 50 m long single mode fiber (SMF), which works as a quantum memory cell \cite{prevedel2011}. We use a free-space electro-optic modulator (EOM) (Qioptiq, LM0202 PHAS) on Bob's side to set the measurement basis, which is triggered by the signal from Alice (connected by FG). Phase compensation (PC) crystals compensate for the birefringent effect. The performance quality of the quantum memory cell and EOM in the absence of a signal from Alice is characterized using quantum process tomograph \cite{chuang1997,poyatos1997,oBrien2004}, with a resulting experimental fidelity of about $0.9802\pm0.0057$ \cite{si}. The state of the photon on Bob's side is analyzed by a QWP, HWP and a polarization beam splitter (PBS). The results detected by detector $D1$ are denoted as $P_{+}$ and $P_{-}$ (successful probabilities) and the results detected by $D2$ are denoted by $P_{+}'$ and $P_{-}'$ (error probabilities).

\begin{figure}
\centering
\includegraphics[width=0.4\textwidth]{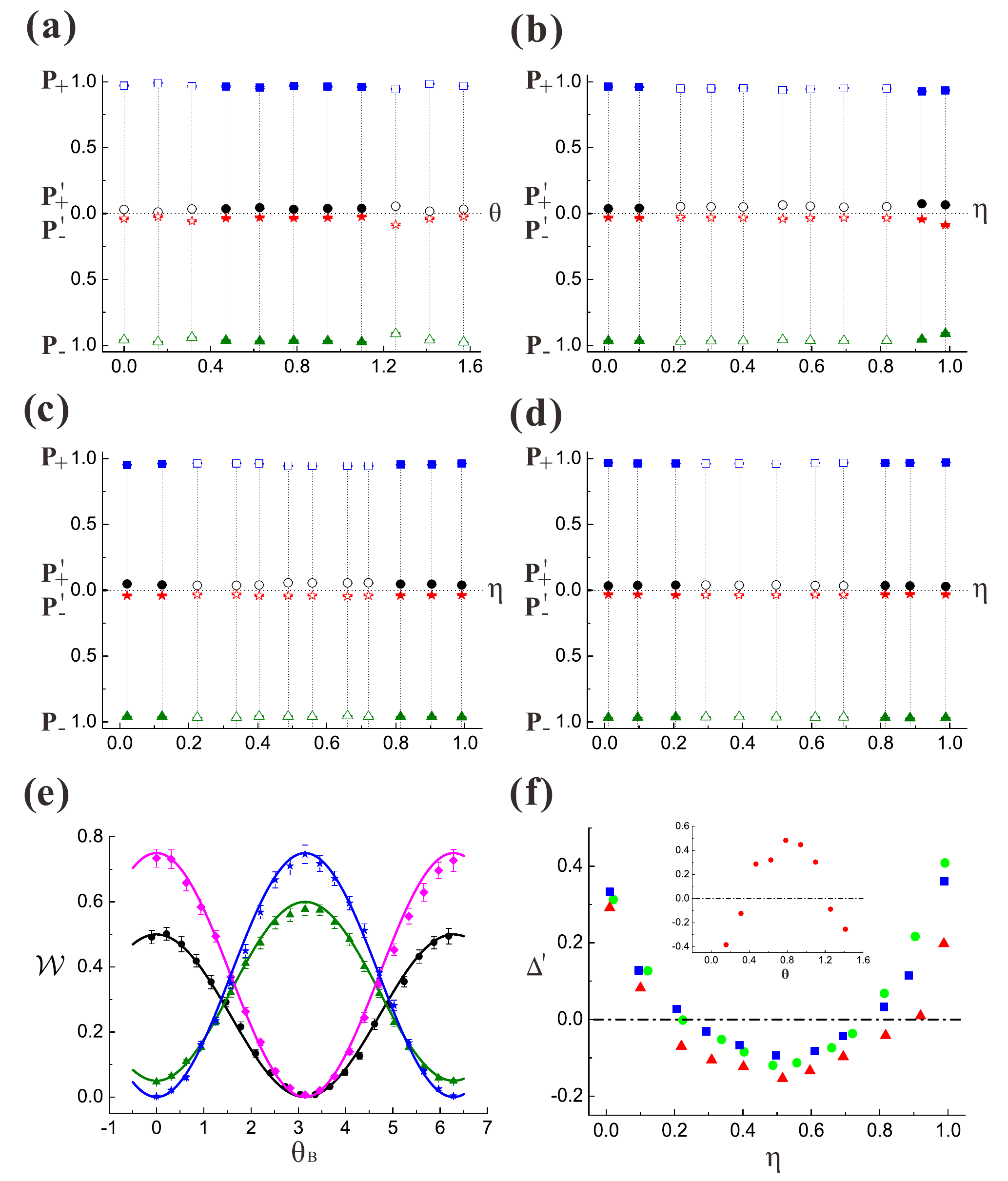}
\caption{{\bf Experimental results for $\rho_1$.} (a)-(d) show the detected probabilities of the NCS on Bob's side. The hollow points represent the states are not steerable in the case of two measurement settings based on the values of $\Delta'$ which are shown in (f), while the solid ones mean the states are steerable. The blue squares and black circles represent the values of $P_{+}$ and $P_{+}'$, respectively. The green up triangles and red stars represent the values of $P_{-}$ and $P_{-}'$, respectively. (e) The values of $\langle\mathcal{W}\rangle$ as a function of $\theta_{B}$. The black circles, red squares, green up triangles, magenta diamonds and blue stars represent cases with input parameters of $\theta=\pi/4$ and $\eta=1$, $\theta=\pi/4$ and $\eta=0$, $\theta=\pi/3$ and $\eta=0.2$, $\theta=\pi/6$ and $\eta=0$, and $\theta=\pi/3$ and $\eta=1$, respectively. The black, red, green, magenta and blue lines represent the corresponding theoretical predictions. (f) The results for $\Delta'$. The red triangles, blue squares and green circles represent the cases with initial parameters of $\theta=\pi/6$, $\theta=\pi/3$ and $\theta=\pi/4$, respectively. The inset in (f) shows the value of $\Delta'$ as a function of $\theta$. The red circles represent the cases with initial parameters of $\eta=1$. The error bars correspond to the counting statistics.}
\label{result1}
\end{figure}

For the kind of states $\rho_1$, Alice measures along the $x$ direction, which leads to two different pure NCS on Bob's side. The eigenvectors of the projector $\sigma_x$ are $1/\sqrt{2} (|H\rangle+|V\rangle)$ and $1/\sqrt{2} (|H\rangle-|V\rangle)$. The corresponding NCS for Bob will be $\cos{\theta}|H\rangle+\sin{\theta}|V\rangle$ and $\cos{\theta}|H\rangle-\sin{\theta}|V\rangle$, with the detected probabilities denoted as $P_+$ and $P_-$, respectively. When $\theta=0$ or $\theta=\pi/2$, the steering game fails, as the initial states represent separable states (i.e., Bob's two NCS are now both equal to $|H\rangle$ or $|V\rangle$). We first show the experimental results for four initial situations, with $\eta=1$ (different $\theta$), and $\theta=\pi/6$, $\theta=\pi/4$ and $\theta=\pi/3$ (different $\eta$) shown in Fig. \ref{result1}(a)-(d), respectively. For each case, the errors (supported by the LHS model) are low. As a result, the AVN demonstration of the steering game is over if Alice and Bob share an entangled state. To check the result, the measurement direction chosen by Alice is $z$, which is orthogonal to $x$. Bob obtains the maximum value of $\langle\mathcal{W}\rangle$ by scanning angle $\theta_B$ (i.e., $\phi_B$ is zero). Fig. \ref{result1}(e) shows some of the experimental results. The angle $\theta_{B}$ required to obtain the maximal value of $\langle\mathcal{W}\rangle$ depends on the initial conditions. According to the LHS model, the upper bound is $C_{LHS}=(1+|\cos2\theta|)/4$, while the quantum prediction for $\langle\mathcal{W}\rangle_{max}$ is $(1/2+|1/2-\eta|)\cdot\cos^2\theta$ when $\theta\in[0,\pi/4]$ and $\theta_B$ is $0$. When $\theta\in[\pi/4,\pi/2]$, $\langle\mathcal{W}\rangle_{max}$ is obtained as $(1/2+|1/2-\eta|)\cdot\sin^2\theta$ with $\theta_B$ being $\pi$. The value of $\Delta={\langle\mathcal{W}}\rangle_{max}-C_{LHS}$ which should not be larger than $0$ according to the LHS model is shown in the supplementary material\cite{si}. According to the AVN criterion, the steering game is successful when $\Delta >0$ as well as $P_+ =1, P_+'=0$ and $P_-=1, P_-'=0$. However, in practice, $P_{\pm}<1$ and $P_{\pm}'>0$. Then we should check whether the inequation (\ref{exp1}) is violated or not. Fig. \ref{result1}(f) shows the value of $\Delta'$. Taking the noise into consideration, we find that for some states, the steering game fails according to the new criterion. To clarify this fact, the not steerable states are marked by the hollow points while the steerable states are marked by the solid points in Fig. \ref{result1}(a)-(d). Note, when $\theta=0$ or $\theta=\pi/2$ in the case of $\eta=1$, it is obvious that the state could not be steerable and the $\Delta'$ is not shown in the figure as its value is much less than zero.

\begin{figure}
\includegraphics[width=0.4\textwidth]{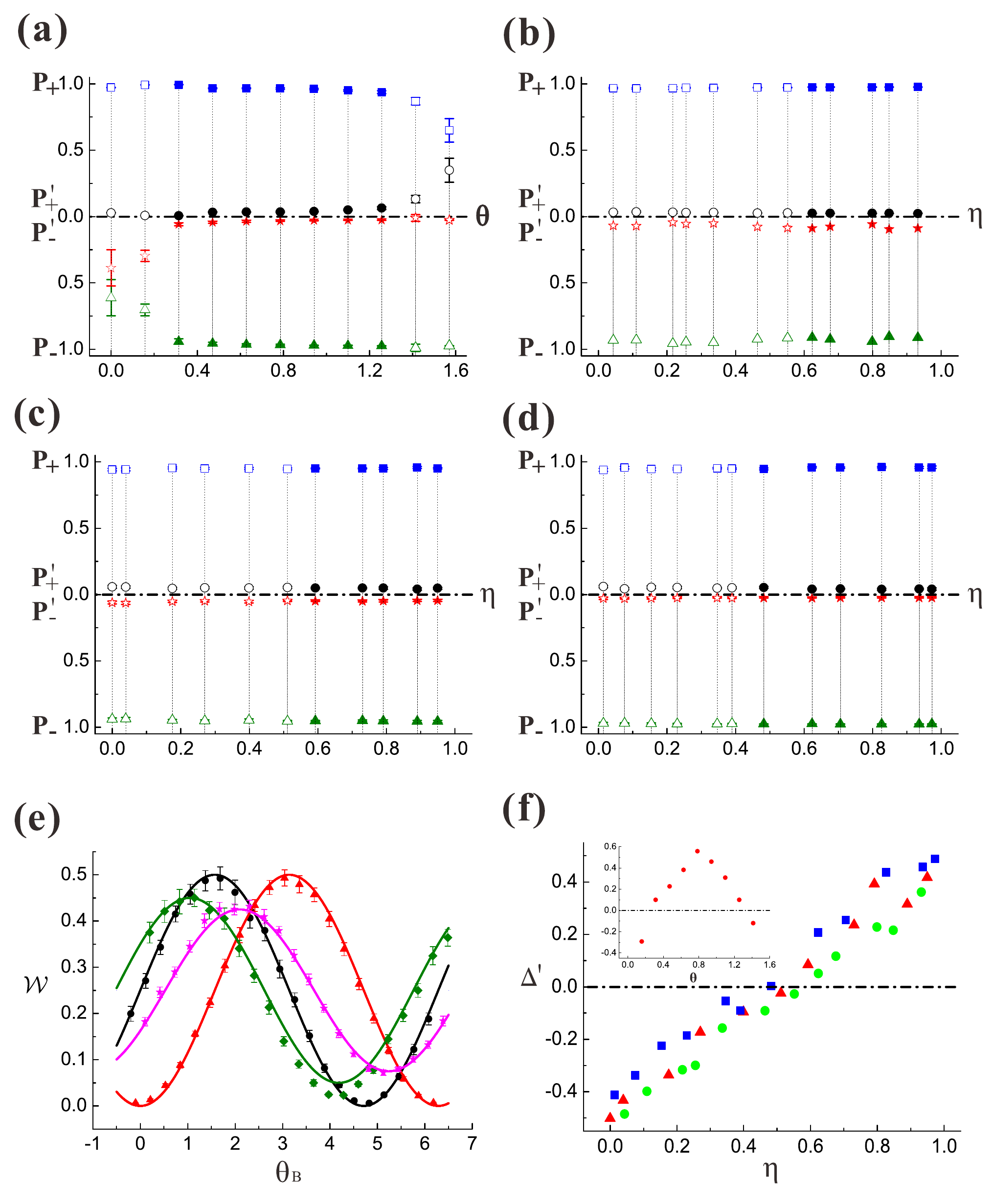}
\caption{{\bf Experimental results for $\rho_2$.} (a)-(d) show the detected probabilities of the NCS on Bob's side. The hollow points represent the states are not steerable in the case of two measurement settings based on the values of $\Delta'$ which are shown in (f), while the solid ones mean the states are steerable. The blue squares and black circles represent the values of $P_{+}$ and $P_{+}'$, respectively. The green up triangles and red stars represent the values of $P_{-}$ and $P_{-}'$, respectively. (e) The values of $\langle\mathcal{W}\rangle$ as a function of $\theta_{B}$. The black circles, red up triangles, green diamonds and magenta stars represent the cases with input parameters of $\theta=\pi/4$ and $\eta=1$, $\theta=\pi/2$ and $\eta=1$, $\theta=\pi/6$ and $\eta=0.8$ and $\theta=\pi/3$ and $\eta=0.7$, respectively. The black, red, green and magenta lines represent the corresponding theoretical predictions. (f) The results for $\Delta'$. The red triangles, blue squares and green circles represent the cases with initial parameters of $\theta=\pi/6$, $\theta=\pi/3$ and $\theta=\pi/4$, respectively. The inset in (f) shows the value of $\Delta'$ as a function of $\theta$. The red circles represent the case with an initial parameter of $\eta=1$. The error bars correspond to the counting statistics.
}
\label{result2}
\end{figure}

We further prepared a second kind of states $\rho_2$ and again implemented the steering game for some states. For these states, Bob's NCS are different pure states if Alice performs the measurement along the $z$ direction. The NCS correspond to $|H\rangle$ and $|V\rangle$ when the eigenvectors of $\sigma_z$ are $|H\rangle$ and $|V\rangle$, respectively. Figs. \ref{result2}(a)-(d) show the experimental probability of a successful detection and the errors for NCS given corresponding initial parameters of (a) $\eta=1$, (b) $\theta=\pi/6$, (c) $\theta=\pi/4$ and (d) $\theta=\pi/3$. When $\eta=1$ and $\theta$ is close to $0$, Bob's NCS $|V\rangle$ almost vanishes. In fact, Bob can isolate the NCS $|H\rangle$, especially when $\theta=0$ (product state). We can see that the error probability approaches the success probability for $P_-$ as $\theta$ approaches $0$, and this is the same case for $P_+$ when $\theta=\pi/2$. Therefore, these two states are clearly not steerable. To check the results, Alice and Bob perform a joint measurement, where the measurement direction on Alice's side is $x$, and Bob scans $\theta_B$ to maximize $\langle\mathcal{W}\rangle$. The experimental result of $\langle\mathcal{W}\rangle$ as a function of $\theta_{B}$ is shown in Fig. \ref{result2}(e). The quantum prediction is $\langle\mathcal{W}\rangle=\eta/2\cdot\cos^2(\theta\pm\theta_B/2)+(1-\eta)/4$, where $\langle\mathcal{W}\rangle_{max}$ is bounded by $C_{LHS}=(1+\eta\cdot|\cos2\theta|)/4$ according to the LHS model. We further show the difference between the results $\Delta$ in the supplementary material \cite{si}. When $\Delta>0$, Bob is convinced that Alice can steer his state in the ideal situation where $P_+ =1, P_+'=0$ and $P_-=1, P_-'=0$. In the experiment, we further check the inequation (\ref{exp1}) to confirm whether the states are steerable or not. The value of $\Delta'$ is shown in Fig. \ref{result2}(f). We can find that some states are verified to be not steerable in the case of two measurement settings. The hollow and solid points represent the not steerable and steerable states, respectively. The states with $\rho_1$ when $\theta=0$ or $\theta=\pi/2$ in the case of $\eta=1$ are product states which are not steerable states and $\Delta'$ is much less than zero which is not shown in the figure. In our experiment, error bars are estimated from standard deviations of the values whose statistical variation are considered to satisfy a Poisson distribution.

In conclusion, we experimentally demonstrated, for the first time, an EPR steering game employing an AVN criterion that strictly follows the practical concept of steering. In our experiment, the AVN criterion was dependent on obtaining two different NCS on Bob's side. To check the results, we measured $\Delta$ for all cases. However, $\Delta$ can be randomly checked if Alice and Bob promise that the initial states are entangled to rule out any cheating from a third party, just like in quantum key distribution~\cite{scarani2009}. Moreover, considering the noise, we develop a new criterion to check the steering. We can therefore verify whether the states are steerable or not depend on the experimental values obtained from the two-setting measurement. Our experimental results provide a particularly strong perspective for understanding EPR steering and has experimental potential applications in the implementation of long-distance quantum information processing ~\cite{marcikic2003,ursin2007,barreiro2008}.

This work is supported by the National Basic Research Program of China (Grants No. 2011CB921200), National Natural Science Foundation of China (Grant Nos. 11274297, 11004185, 61322506, 60921091, 11274289, 11325419, 61327901, 11275182), the Fundamental Research Funds for the Central Universities (Grant Nos. WK 2030020019, WK2470000011), Program for New Century Excellent Talents in University (NCET-12-0508), Science foundation for the excellent PHD thesis (Grant No. 201218) and the CAS. JLC acknowledges the support by National Basic Research Program of China under Grant No. 2012CB921900 and NSF of China (Grant Nos. 11175089, 11475089). This work is also partly supported by National Research Foundation and Ministry of Education, Singapore.

\clearpage
\subsection{Supplementary Material: Experimental demonstration of the Einstein-Podolsky-Rosen steering game based on the All-Versus-Nothing proof}

\subsection{State preparation using the unbalanced Mach-Zehnder interferometer}
In our experiment, the ultraviolet pulses with a wavelength of $400$ nm and a bandwidth about $1.21$ nm pump two type-I BBO crystals to generate entangled photon pairs via spontaneous parametric down conversion (SPDC). And two kinds of unbalanced Mach-Zehnder (UMZ) interferometers are used to prepare the initial states shown in the equation (5) in the main text. A similar method has been used in our previous experiment \cite{jsx2013}. Here, we explain how the UMZ interferometers work.

The UMZ interferometer {\bf (a)} in Fig. 2 in the main text is a kind of polarization-independent interferometer including two polarization-independent beam splitters (BS). The polarization state is separated into two parts by the first BS. We can then conveniently apply suitable single qubit unitary operation to the photon passing through the long arm or short arm, and obtain the corresponding two-photon state. The relative amplitude of the two parts can also be conveniently adjusted by inserting a shutter into one of the two arms. For example, we can implement a half-wave plate (HWP) with the angle setting at $45^\circ$ in the long arm (i.e., HWP2 in the main text) which rotates $|H\rangle$ to $|V\rangle$ and $|V\rangle$ to $|H\rangle$. For the initial input state $|\Psi (\theta) \rangle = \cos {\theta}|HH\rangle + \sin{\theta}|VV\rangle$ where $\theta$ is the angle of the HWP1 in the main text, the state in the long arm becomes $|\Phi (\theta)\rangle =\cos{\theta}|VH\rangle+\sin{\theta}|HV\rangle$. We do nothing in the short arm and the state remains $|\Psi (\theta)\rangle$. These two parts then combine again by the second BS. The time difference between the long arm and short arm is much larger than the coherence time of the photons (about 0.711 ps), which is traced over during the detection. The state prepared by Alice can then be written as
\begin{equation}\label{stap}
\rho_1=\eta|\Psi (\theta)\rangle \langle\Psi (\theta)|+(1-\eta)|\Phi (\theta)\rangle \langle\Phi (\theta)| \tag{S1}
\end{equation}
where $\eta$ represents the relative amplitude of these two arms.

To prepare $\rho_2$, a polarization-dependent UMZ interferometer {\bf (b)} which consists of two polarization beam displacers (BD) is inserted into the long arm of the UMZ interferometer {\bf (a)}. In such case, the HWP (i.e., HWP2 in the main text) is set to be $0^\circ$ and the state remains $|\Psi (\theta)\rangle$. The UMZ interferometer {\bf(b)} further separates the parts of $|HH\rangle$ and $|VV\rangle$. The relative amplitude between these two parts can be adjusted conveniently by inserting a shutter into one of the arms (in this experiment, the amplitudes are adjusted to be equal to each other). There are time differences when $|HH\rangle$ and $|VV\rangle$ combine again by the second BD. To further completely distinguish the arrival time of $|HH\rangle$ and $|VV\rangle$, we implement a thick quartz plate, in which the horizontal and vertical polarization components have different velocities. After these two UMZ interferometers, the time differences among the three parts of $|\Psi (\theta)\rangle$, $|HH\rangle$ and $|VV\rangle$ are all larger than the coherence time of the photons, which are traced over during the detection. The final state for $\rho_2$ reads as
\begin{equation}\label{stap1}
\rho_2=\eta|\Psi (\theta)\rangle \langle \Psi (\theta)|+\frac{1-\eta}{2}(|HH\rangle \langle HH|+|VV\rangle \langle VV|). \tag{S2}
\end{equation}

\subsection{Time delay in the classical channel}
The $50$ m single-mode fiber has a refractive index of about $1.5$, and provides a time delay of about $250$ ns for Bob to respond to Alice's signals. The time delays in the classical channel includes the rising time of the signals from the single-photon detector ($\sim$10 ns), the function generator ($\sim$5 ns), the driver of the electro-optic modulator (EOM) ($\sim$22 ns), the response time of the function generator ($\sim$15 ns) and the driver ($\sim$150 ns). Therefore, the minimal transmission time of the signal from Alice to Bob is less than the amount of time the photon could be stored in quantum memory. To count photons within the coincidence window ($\sim$3 ns), a coaxial-cable with a length of about $45$ m is used to transmit the electric signals detected by Alice.

\subsection{Quantum process tomography on Bob's side}

\begin{figure}
\centering
\includegraphics[width=0.45\textwidth]{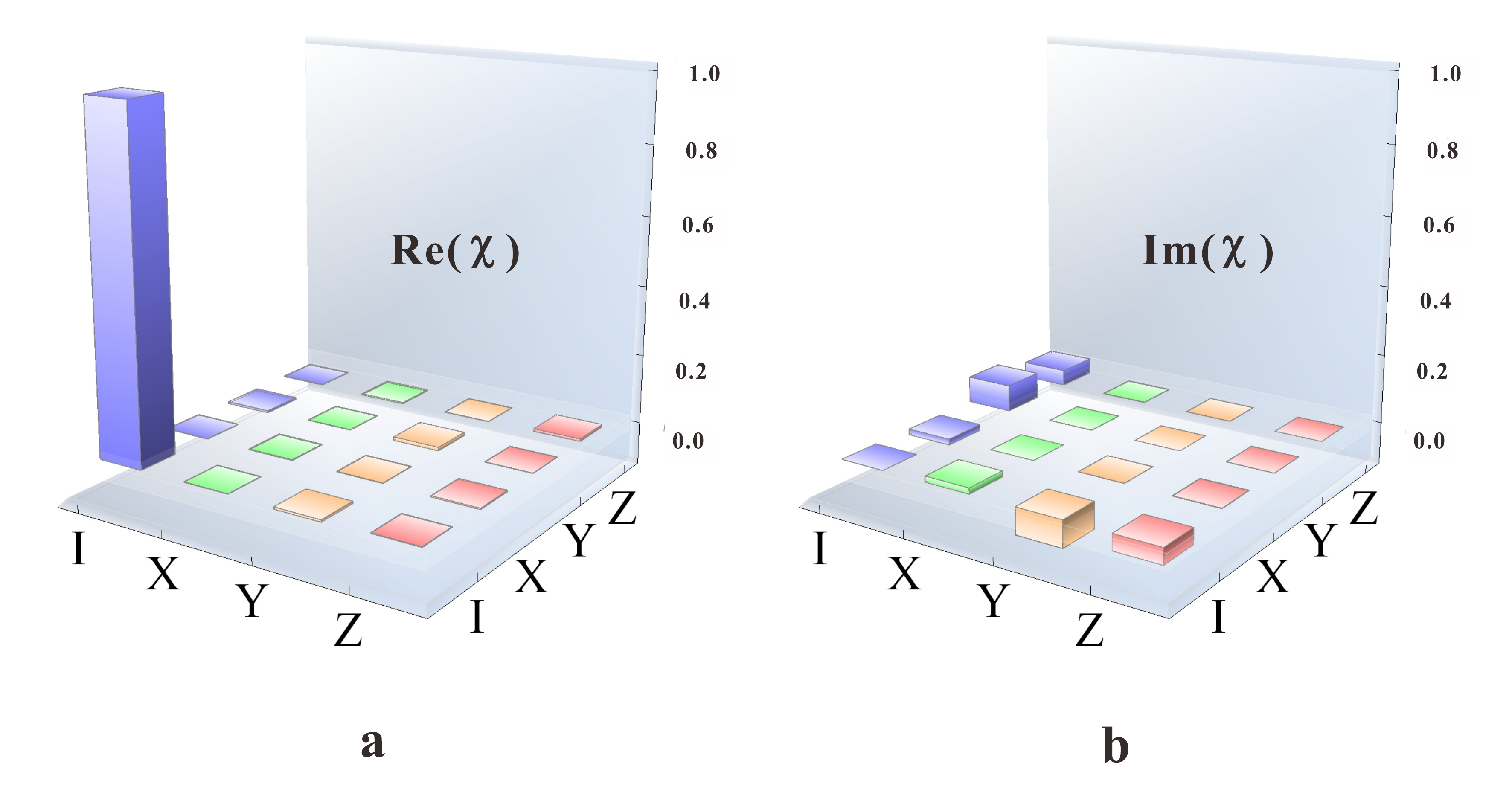}\\
\renewcommand{\thefigure}{S1}
\caption{Experimental results of the density matrix $\chi$ on Bob's side; {\bf{(a)}} and {\bf{(b)}} represent the real and imaginary parts, respectively.}
\label{tomo}
\end{figure}

The experimental setup in Bob's side includes a $50$ m single mode fiber (SMF) which serves as a quantum memory and a free-space electro-optic modulator (EOM) which is used to respond to the high speed electrical signal sent from Alice and changed the measurement settings. Two HWPs are used to compensate the basis rotation in the single-mode fiber. We use a crystal to compensate for birefringence caused by the EOM. To characterize the performance of these components, we employed a quantum process tomography approach~\cite{chuang1997,poyatos1997,oBrien2004}. The operator basis ($E_i$) was chosen to include the identity operation ($I$) and the three Pauli operators ($X$, $Y$ and $Z$). The operator of the combined unit can be written as
\begin{equation}\label{qpt}
\varepsilon_\rho=\sum_{ij} \chi_{ij}E_i\rho E_j^\dag \tag{S3}
\end{equation}where $\chi$ is a complex matrix describing the process $\varepsilon_\rho$. In our experiment, we estimated $\chi$ using the maximum-likelihood procedure \cite{oBrien2004} with the results shown in Fig.~\ref{tomo}. The fidelity calculated from $(Tr\sqrt{\sqrt{\chi}\chi_{ideal}\sqrt{\chi}})^2$ with $\chi_{ideal}=I$ is about $98.02\%\pm 0.57\%$.

\subsection{More experimental results}
\subsubsection{Results for states $\rho_1$}

\begin{figure}
\centering
\includegraphics[width=0.48\textwidth]{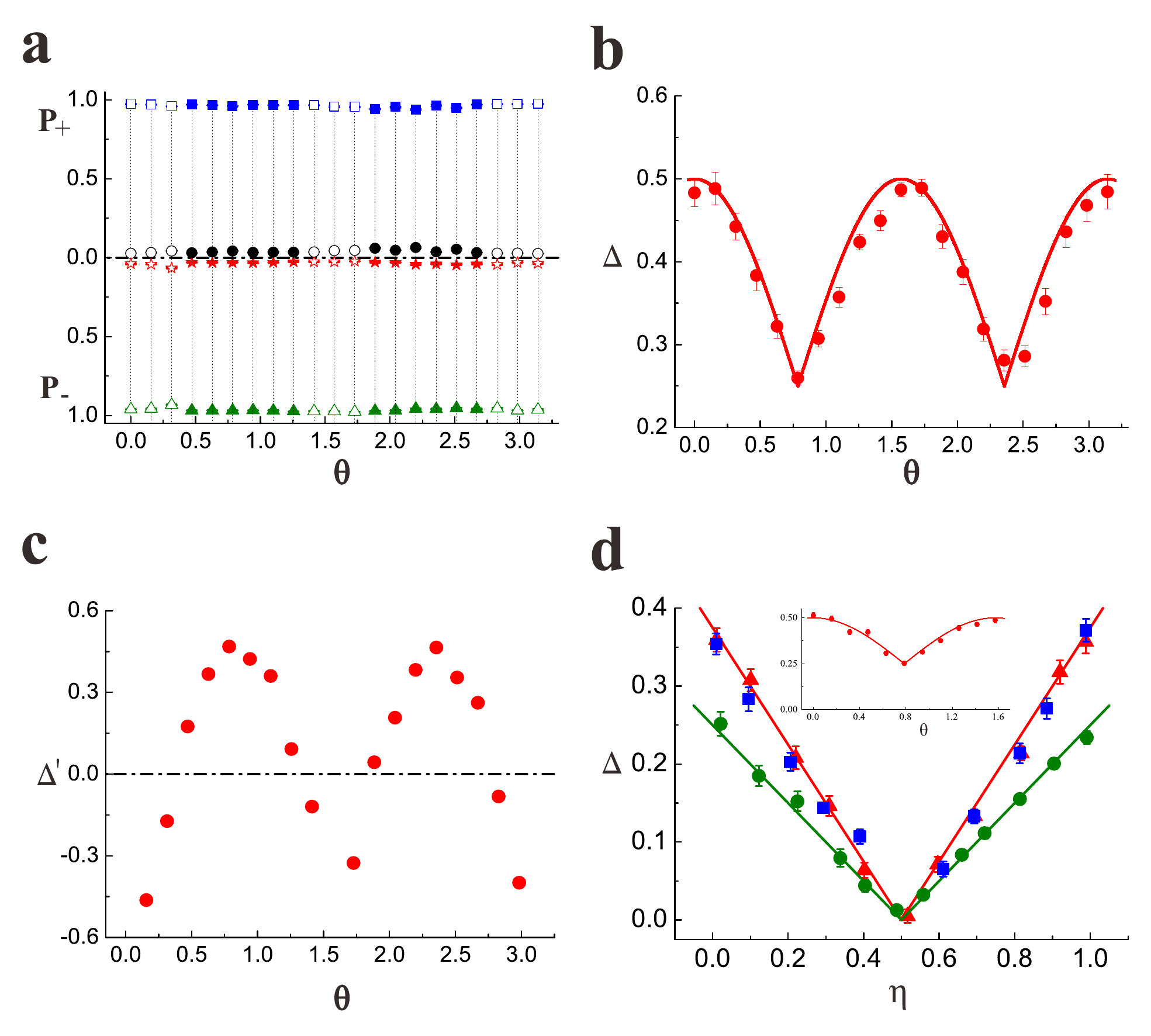}\\
\renewcommand{\thefigure}{S2}
\caption{More experimental Results for $\rho_1$. {\bf{a-c}} $\eta=0$. {\bf{a}} The detected probabilities of the NCS on Bob's side. The blue squares and black circles represent the values of $P_{+}$ and $P_{+}'$, respectively. The green triangles and red stars represent the values of $P_{-}$ and $P_{-}'$, respectively. And the hollow and solid points represent the not steerable and steerable states, respectively. {\bf{b}} The results of $\Delta$. The red dots represent the experimental results, while the red solid line represents the corresponding theoretical prediction. {\bf{c}} The value of $\Delta'$ which is developed with the experimental noise. {\bf{d}} The values of $\Delta$ of other states of $\rho_1$. The red triangles, blue squares and green circles represent the cases with initial parameters of $\theta=\pi/6$, $\theta=\pi/3$ and $\theta=\pi/4$, respectively. The red, blue and green lines represent the corresponding theoretical predictions. Note that the red and blue lines completely overlap so only the red line can be seen. The inset in (d) shows the value of $\Delta$ as a function of $\theta$ when $\eta=1$. The line represents the corresponding theoretical predictions.}
\label{addresult}
\end{figure}

Figure.~\ref{addresult} shows the experimental results for the initial state of $\cos \theta |HV\rangle +\sin \theta |VH\rangle$ ($\eta=0$ in $\rho_{1}$). The detected probabilities are shown in Fig.~\ref{addresult}(a). In the figure, the blue squares and black circles represent the values of $P_{+}$ and $P_{+}'$, respectively while the green triangles and red stars represent the values of $P_{-}$ and $P_{-}'$, respectively. The values of $P_{+}'$ and $P_{-}'$ are small, and correspond to the error in detecting the pure normalized conditional states (NCS) on Bob's side. Fig.~\ref{addresult}(b) shows the value of $\Delta$. The red dots represent the experimental results with the black solid line representing the corresponding theoretical prediction. While Fig. ~\ref{addresult}(c) shows the values of $\Delta'$ which involves the noise in the experiment. Note, when $\theta=0,\pi/2$ and $\pi$, the state is not steerable obviously and  $\Delta'$ is not shown in the figure. We also measure $\Delta$ of other states of $\rho_1$ and the values are shown in Fig.~\ref{addresult}(d). We can see that $\Delta>0$ as shown in Fig. ~\ref{addresult}(b) and (d). According to the new criterion, the states with $\Delta'\leq 0$ are checked to be not steerable.

\subsubsection{Results for states $\rho_2$}

\begin{figure}
\centering
\includegraphics[width=0.46\textwidth]{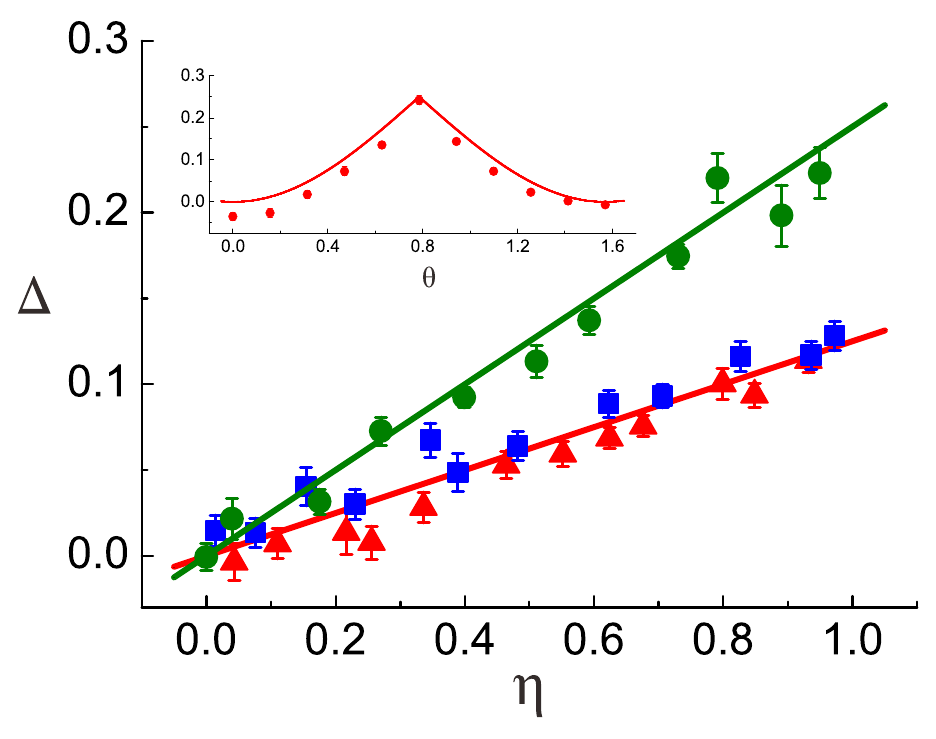}\\
\renewcommand{\thefigure}{S3}
\caption{More experimental Results for $\rho_2$. The results for $\Delta$. The red triangles, blue squares and green circles represent the cases with initial parameters of $\theta=\pi/6$, $\theta=\pi/3$ and $\theta=\pi/4$, respectively. The red, blue and green lines represent the corresponding theoretical predictions. Note that the red and blue lines completely overlap so only the red line can be seen. The inset shows the value of $\Delta$ as a function of $\theta$. The red circles represent the case with an initial parameter of $\eta=1$, where the red line represent the corresponding theoretical prediction.}
\label{addresult2}
\end{figure}

Fig.~\ref{addresult2} shows the values of $\Delta$ for the states of $\rho_2$ with the initial parameters of $\theta=\pi/6, \pi/3$ and $\pi/4$. Although $\Delta$ is larger than zero, some states are checked to be not steerable according to the new criterion with experimental errors taking into consideration.

\subsection{The new criterion dealing with the noise}

The All-Versus-Nothing (AVN) criterion demands that two different pure states are received by Bob. However, in practice, this result can not be achieved. Noise and measurement statistics mean that one can not be sure that the measured state is indeed pure. We develop a theoretical method to deal with the experimental errors involving in the application of the AVN criterion. Before we present this more general criterion, let us make some definitions and prove some lemmas first.

\begin{figure}
\centering
\includegraphics[width=0.46\textwidth]{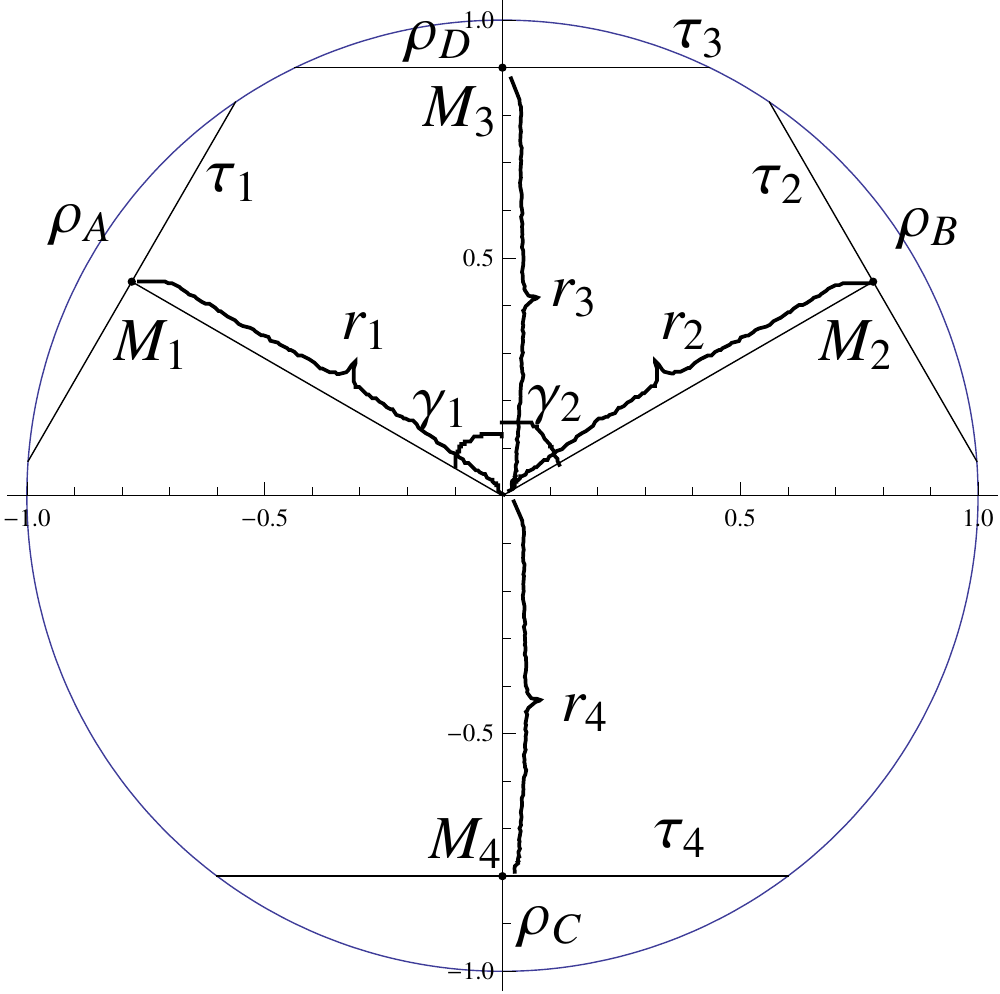}\\
\renewcommand{\thefigure}{S4}
\caption{Experiment results show on Bob's Bloch sphere. $\rho_{A},\rho_{B},\rho_{C}$ and $\rho_{D}$ stand for $\rho_{1}^{\hat{x}},\rho_{0}^{\hat{x}},\rho_{0}^{\hat{z}}$ and $\rho_{1}^{\hat{z}}$ respectively. $M_i$ ($i=1,2,3,4$) are given by the experiment result. The length of $M_i$ are $r_i$, which satisfy $r_1=1-2 \langle W_1 \rangle$, $r_2=1-2 \langle W_2 \rangle$, $r_3=2 {\langle W_3 \rangle}-1$ and $r_4=2 \langle W_4 \rangle-1$. $\gamma_1$ are the angle between $M_1$ and z-axis, similar is the $\gamma_2$, they are determined by the measurement direction of Bob. Tangents $\tau_i$ is perpendicular to $M_i$ and pass it.}\label{figure1}
\end{figure}

As show in Fig.(\ref{figure1}), $\rho_{a}^{\hat{A}}$ ($\hat{A}\in\{\hat{x},\hat{z}\}$,$a\in\{0,1\}$) is the conditional state at Bob's side after Alice's measurement. For the sake of convenience, let us name $\rho_{1}^{\hat{x}},\rho_{0}^{\hat{x}},\rho_{0}^{\hat{z}}$ and $\rho_{1}^{\hat{z}}$ as $\rho_{A},\rho_{B},\rho_{C}$ and $\rho_{D}$. Their Bloch vectors are $A,B,C$ and $D$. $M_i$ ($i=1,2,3,4$) are given by the experiment results at Bob's side, which intuitively show Bob's measurement results. The length of $M_i$ are $r_i$, $r_1=1-2 \langle W_1 \rangle$, $r_2=1-2 \langle W_2 \rangle$, $r_3=2 {\langle W_3 \rangle}-1$ and $r_4=2 \langle W_4 \rangle-1$. The direction of $M_1, M_2$ are determined by $\gamma_1,\gamma_2$ which are given by Bob's measurement directions. $M_3, M_4$ are on z-axis. Tangents $\tau_i$ is perpendicular to $M_i$ and pass it. With the experiment data $\langle W_i \rangle$ and $P_j$($j=A,B,C,D$), Bob doesn't know where the conditional states $A,B,C,D$ are, but he can confirm they are land on $\tau_i$ respectively. Based on this constraint we can find a criterion to demonstrate steering nonlocality. Let us consider the ideal case first, suppose $\gamma_1=\gamma_2=\gamma$, $r_1=r_2=r_{12}$ and $M_3, M_4$ are on z-axis. Now we present the lemmas which are used in the derivation of the criterion.

\noindent \emph{Definition 1}. A local hidden state model (LHSM) of Alice to steer Bob's states is a quantum state ensemble $\{\wp_{\xi}\rho_{\xi}\}$
which gives:
\begin{equation}\label{LHSM}
\tilde{\rho}^{\hat{A}}_{a}=\sum_{\xi} \wp(a|\hat{A},\xi)\wp_{\xi}\rho_{\xi} \tag{S4}
\end{equation}
where $\tilde{\rho}^{\hat{A}}_{a}$ is the conditional states of Bob after Alice measures $\hat{A}$ and gets result $a\in\{0,1\}$, the tilde here
denotes  this state is unnormalized and its norm is $P^{\hat{A}}_{a}$, the probability associated with the output $a$.
$\rho_{\xi}$ is the ``hidden state'' at Bob's side specified by the parameter $\xi$ and $\wp_{\xi}$ is its weight in the ensemble, $\wp(a|\hat{A},\xi)$ is the probability associated with a stochastic map from $\xi$ to $a$ which satisfies positivity.

\noindent \emph{Definition 2}. A deterministic local hidden state model (dLHSM) is a LHSM which satisfies
 $\wp(a|\hat{A},\xi) \in \{0,1\}, \forall \hat{A}, \xi, a$.

\noindent \emph{Lemma 1}.--- For any given two qubit state $\rho_{AB}$, if there is a LHSM for $\rho_{AB}$ then there is a dLHSM for $\rho_{AB}.$ The proof could be found in Ref. \cite{AVNEx}.

\noindent \emph{Lemma 2.---} For a dLHSM, Eq.~(\ref{LHSM}) can be rewritten as $P^{\hat{A}}_{a} \rho^{\hat{A}}_{a}=\sum_{\xi \in H^{\hat{A}}_{a}} \wp_{\xi} \rho_{\xi}$. Here, $H^{\hat{A}}_{a}$ stands for the set of all hidden states which have a contribution to $\rho^{{\hat{A}}}_{a}$, $H^{\hat{A}}_{a}=\{\xi\; |\; \wp(a|\hat{A},\xi)=1 \}$. Lemma 2 says this equality holds if and only if the following equalities hold:
\begin{equation}\label{MassCenter}
\left\{ \begin{aligned}
         P^{{\hat{A}}}_{a} &= \sum_{\xi \in H^{\hat{A}}_{a}} \wp_{\xi} \\
         P^{{\hat{A}}}_{a} \; \overrightarrow{r_{a}}^{{\hat{A}}} &= \sum_{\xi \in H^{\hat{A}}_{a}} \wp_{\xi} \; \overrightarrow{r_{\xi}}
        \end{aligned} \right. \tag{S5}
\end{equation}
where $\overrightarrow{r_{a}}^{{\hat{A}}}$and $\overrightarrow{r_{\xi}}$ stand for the Bloch vectors of $\rho^{{\hat{A}}}_{a}$ and $\rho_{\xi}$ respectively. See the proof in \cite{AVNEx}.

Notice that Eqs. \eqref{MassCenter} is similar to the definition of the center of mass if we treat the probabilities $\wp_{\xi}$ and $P^{\hat{A}}_{a}$ as masses and Bloch vectors ( $\overrightarrow{r_{\xi}}$ and $\overrightarrow{r_{a}}^{{\hat{A}}}$) as position vectors of various masses. Lemma 2 shows the task of finding a dLHSM description for a state $\rho^{\hat{A}}_{a}$ with probability $P^{\hat{A}}_{a}$ is equivalent to find a distribution of masses in the Bloch sphere whose total mass is $P^{\hat{A}}_{a}$ and whose center of mass is located at $\overrightarrow{r_{a}}^{{\hat{A}}}$. Those two requirements give constraints to the possibility of finding a dLHSM, for different measurement settings. If we cannot find a dLHSM for $\rho_{AB}$, lemma 1 shows that we cannot find a LHSM neither, thus affirming the steerability of $\rho_{AB}$.

\begin{figure}[h]
\includegraphics[width=0.46\textwidth]{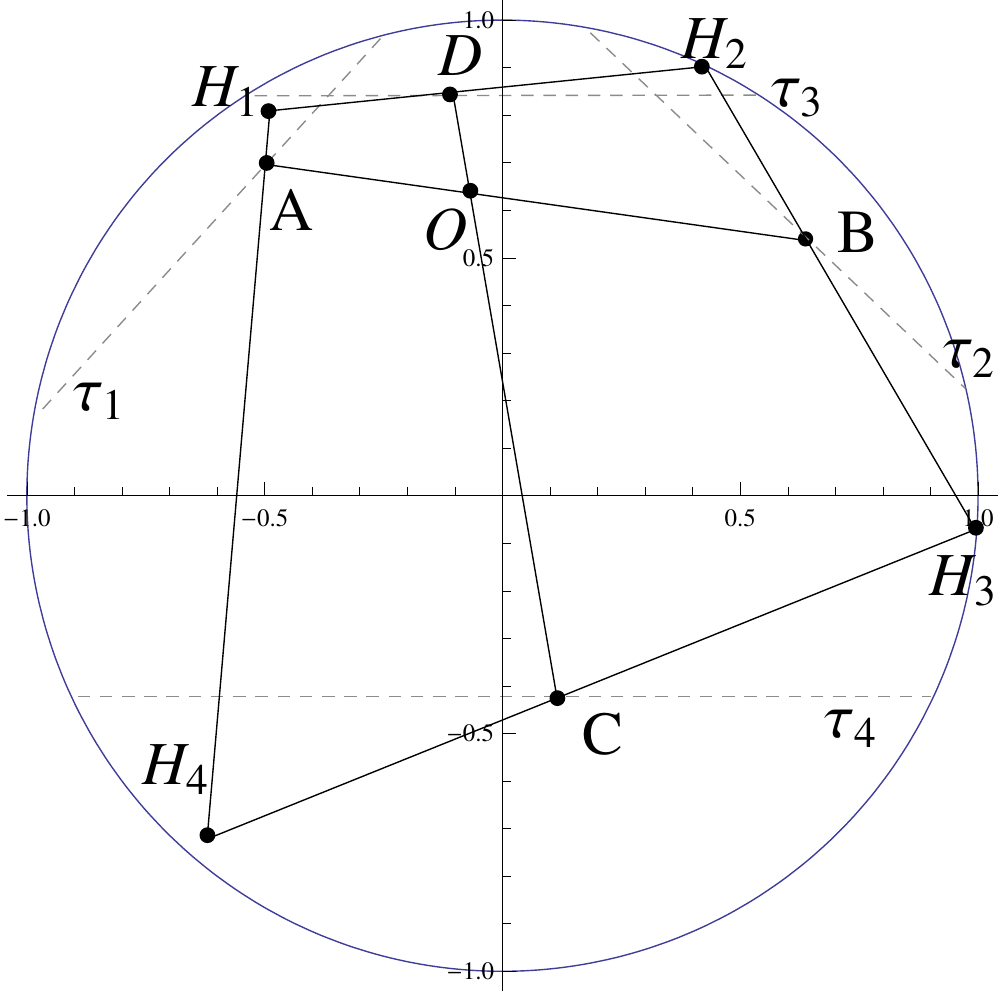}\\
\renewcommand{\thefigure}{S5}
\caption{Four hidden states to simulate the conditional states. $H_i$ ($i=1,2,3,4$) are the hidden states, $A,B,C$ and $D$ stand for the conditional states. As Eqs. \eqref{LHSM4} show, $H_1, H_4$ simulate $A$, $H_2, H_3$ simulate $B$, $H_3, H_4$ simulate $C$ and $H_1, H_2$ simulate $D$.}\label{figure2}
\end{figure}

\noindent \emph{Lemma 3.---} For any given two-qubit state $\rho_{AB}$ in a N-setting protocol, if there is a LHSM for $\rho_{AB}$, then there is a dLHSM with the number of hidden states no larger than $2^N$. In the two setting case we consider
here, as show in Fig.(\ref{figure2}), lemma 3 says 4 hidden states $H_i$ ($i=1,2,3,4$) is enough, their probability are $P_i$. Eqs. \eqref{MassCenter} could rewriter as:

\begin{equation}\label{LHSM4}
\left\{ \begin{aligned}
         P_{A} &= P_1+P_4\texttt{;} \;\;P_A A=P_1 H_1 + P_4 H_4 \\
         P_{B} &= P_2+P_3\texttt{;} \;\;P_B B=P_2 H_2 + P_3 H_3 \\
         P_{C} &= P_3+P_4\texttt{;} \;\;P_C C=P_3 H_3 + P_4 H_4 \\
         P_{D} &= P_1+P_2\texttt{;} \;\;P_D D=P_1 H_1 + P_2 H_2
        \end{aligned} \right. \tag{S6}
\end{equation}

Lemma 3 tells us if there isn't an ensemble $\{P_i H_i\}$ which makes Eqs. \eqref{LHSM4} hold then the state we discussed is
steerable. The proof could find in Ref. \cite{AVNEx}.

\noindent \emph{Lemma 4.---} If there is an ensemble $\{P_i H_i\}$ which makes Eqs. \eqref{LHSM4} hold then there is $\{P'_i H'_i\}$ with $H'_i$ on xoz-plan makes Eqs. \eqref{LHSM4} be satisfied.

\noindent \emph{Proof of Lemma 4.---} Suppose $\{P_i H_i\}$ makes Eqs. \eqref{LHSM4} hold, we could construct $\{P'_i H'_i\}$ as follows: $P'_i=P_i$, $H'_i=\{H_{ix},0,H_{iz}\}$. Here $H_{ix}, H_{iz}$ stands for the x-component and z-component of $H_i$. Notice $A,B,C$ and $D$ are on xoz-plane, by direct calculation we could see $\{P'_i H'_i\}$ is an appropriate ensemble which satisfy Eqs. \eqref{LHSM4}. From now on we could always assume the hidden states are on the xoz-plane, lemma 4 tales us if there isn't such states on the xoz-plane satisfy Eqs. \eqref{LHSM4} then steering is demonstrate.

\noindent \emph{Lemma 5.---} The experiment result doesn't give the exact location of $A,B,C,D$ but only restrict them on $\tau_i$. Thus to check whether there is $\{P_i H_i\}$ satisfy Eqs. \eqref{LHSM4}, $A,B,C,D$ are also variables restrict on $\tau_i$. lemma 5 says if Eqs. \eqref{LHSM4} could be hold by variables $\{P_i H_i\}$, $A,B,C,D$ and $P_j$ ($j=A,B,C,D$) then there is another group of variables  $\{P'_i H'_i\}$, $A',B',C',D'$ and $P_{j'}$ satisfy Eqs. \eqref{LHSM4} which has the followed restriction on itself:
\begin{equation}\label{LHSM5}
\left\{ \begin{aligned}
         &P_{A'} = P_{B'}=\frac{1}{2}; \;\; P_{C'} = P_{C};\;\;P_{D'} = P_{D}   \\
         &\{A'_x,A'_z\} = \{-B'_x,B'_z\};\;\;C'=M_4=\{0,C_z\}   \\
         &D'=M_3=\{0,D_z\} \\
         &\{H'_{1x},H'_{1z}\} = \{-H'_{2x},H'_{2z}\}    \\
         &\{H'_{3x},H'_{3z}\} = \{-H_{4x},H_{4z}\}\\
         &P'_{1} = P'_{2}=\frac{P_D}{2};\;\;P'_{3} = P'_{4}=\frac{P_C}{2}
         \end{aligned} \right. \tag{S7}
\end{equation}
\noindent \emph{Proof of Lemma 5.---} Let us prove it by constructing the required $\{P'_i H'_i\}$, $A',B',C',D'$ and $P'_j$.
\begin{equation}\label{LHSM6}
\left\{ \begin{aligned}
         &A' = P_{A} A +P_{B} \{-B_x,B_z\}  \\
         &B' = P_{B} B +P_{A} \{-A_x,A_z\}  \\
         &C'=M_4;\;\;D'=M_3  \\
         &P_{A'}=P_{B'}=\frac{1}{2};\;\;\;P_{C'} = P_{C};\;\;P_{D'} = P_{D}\\
         &H'_1=\frac{P_1}{P_D}H_1+\frac{P_2}{P_D} \{-H_{2x},H_{2z}\}  \\
         &H'_2=\frac{P_2}{P_D}H_2+\frac{P_1}{P_D} \{-H_{1x},H_{1z}\}\\
         &H'_3=\frac{P_3}{P_C}H_3+\frac{P_4}{P_C} \{-H_{4x},H_{4z}\}  \\
         &H'_4=\frac{P_4}{P_C}H_4+\frac{P_3}{P_C} \{-H_{3x},H_{3z}\}\\
         &P'_{1} = P'_{2}=\frac{P_D}{2};\;\;P'_{3} = P'_{4}=\frac{P_C}{2}
         \end{aligned} \right. \tag{S8}
\end{equation}

\begin{figure}[h]
\includegraphics[width=0.46\textwidth]{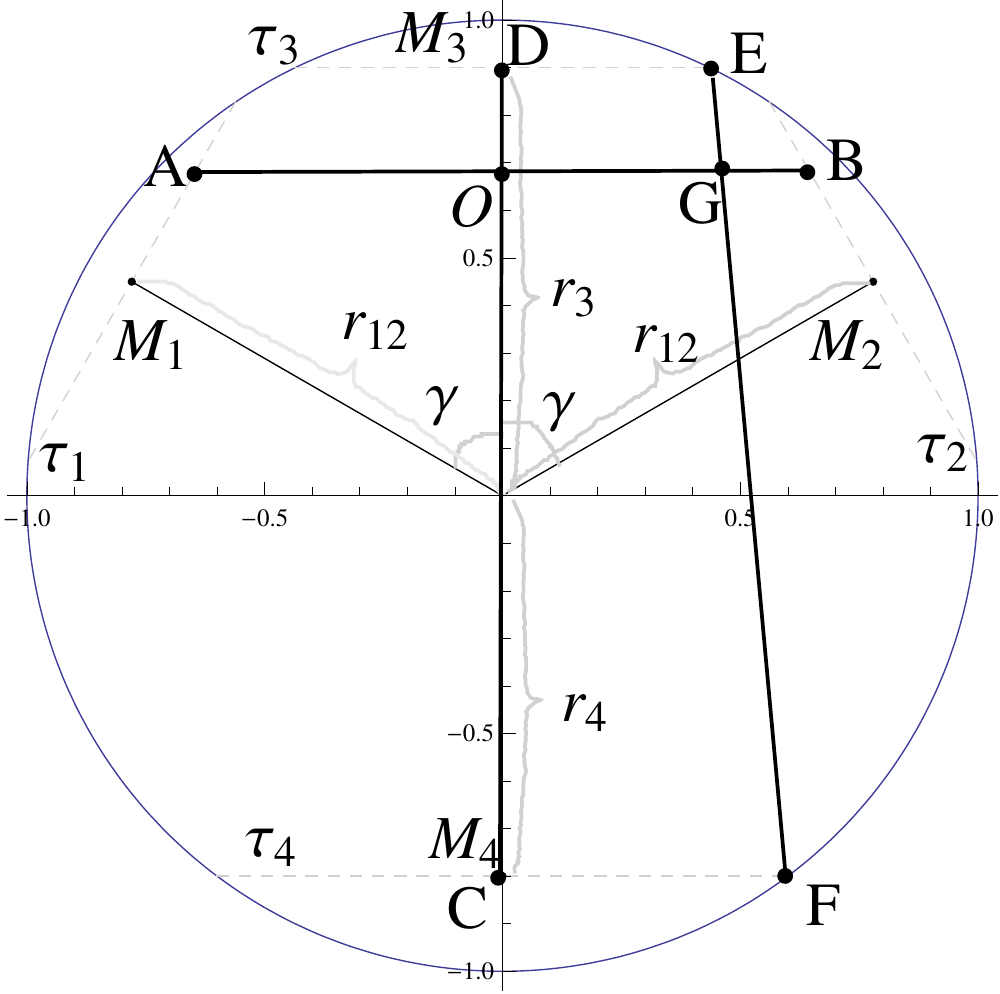}\\
\renewcommand{\thefigure}{S6}
\caption{The steering criterion show on Bob's Bloch sphere. $A,B$ are symmetry to z-axis, $D=M_3$, $C=M_4$ and $O$ stands for $\rho_{Bob}=\textbf{Tr}_A[\rho_{AB}]$. $E=\{\sqrt{1-M_{3z}^2},M_{3z}\}$, $F=\{\sqrt{1-M_{4z}^2},M_{4z}\}$, $G$ is the point where $EF$ cuts $AB$.}\label{figure3}
\end{figure}

To prove this lemma we need check two requires, the first one is $A',B',C',D'$ is on $\tau_i$ respectively. Take $A'$ for example, notice that $A$ and $\{-B_x,B_z\}$ are on $\tau_1$, so $A'$ as their convex combination also on $\tau_1$. The others could be checked similarly. Next we need to check it satisfy Eqs. \eqref{LHSM4}. Take the first equation for example, according to the definition we get $P'_1+P'_4=\frac{P_D}{2}+\frac{P_C}{2}=\frac{1}{2}$ thus $P'_1+P'_4=P_{A'}$. And
\begin{equation}
\begin{split}\label{LHSM7}
P'_1 H'_1 + P'_4 H'_4 & =\frac{P_D}{2} \frac{P_1}{P_D} \{H_{1x},H_{1z}\} + \\
 &\frac{P_D}{2} \frac{P_2}{P_D} \{-H_{2x},H_{2z}\} +\frac{P_C}{2} \frac{P_4}{P_C} \{H_{4x},H_{4z}\} \\
 &+\frac{P_C}{2} \frac{P_3}{P_C} \{-H_{3x},H_{3z}\}\\
&=\frac{1}{2}\{P_1 H_{1x}+P_4 H_{4x},\;\;P_1 H_{1z}+P_4 H_{4z}\}+ \\
&\frac{1}{2}\{-P_2 H_{2x}-P_3 H_{3x}, \;\;P_2 H_{2z}+P_3 H_{3z}\}\\
&=\frac{1}{2}P_{A} A +\frac{1}{2}P_{B} \{-B_x,B_z\}=P_{A'} A'
\end{split} \tag{S9}
\end{equation}
Similarly, the other equations could also be checked. According to the definition Eqs. \eqref{LHSM6}, the requires Eqs. \eqref{LHSM5} could also easily be checked. Thus this lemma has been proved. It tells us if there isn't a LHSM $\{P'_i,H'_i \}$ on an isosceles trapezoid to represent the symmetrical conditional state $A',B',C',D'$ then there is no LHSM for the original $A,B,C,D$ thus the experimental results demonstrate EPR steering.

\subsubsection{Criterion of Steering.}
 As show in Fig.(\ref{figure3}), $A,B$ are symmetry to z-axis, $D=M_3$, $C=M_4$ and $O$ stands for $\rho_{Bob}=\textbf{Tr}_A[\rho_{AB}]$. Define $E=\{\sqrt{1-M_{3z}^2},M_{3z}\}$, $F=\{\sqrt{1-M_{4z}^2},M_{4z}\}$, $G$ is the point where $EF$ cuts $AB$. This criterion says if $|OB|-|OG|>0$ then there is no LHSM satisfy Eqs. \eqref{LHSM4}. Thus the state $\rho_{AB}$ we discussed is steerable.

\noindent \emph{Proof of the Criterion.---} The proof is quite intuitive, see Fig.(\ref{figure3}), lemma 5 tales us if exists LHSM satisfy the experimental result then it always exists a LHSM form as an isosceles trapezoid. Notice that $H_1,H_2$ are on $\tau_3$ and $H_3,H_4$ are on $\tau_4$ thus if $|OB|>|OG|$ then the requirement $P_B B=P_2 H_2 + P_3 H_3$ can not be satisfied. Thus the state we discussed is steerable. While, if $|OB|\leq |OG|$, it is easy to see we could find a LHSM to simulate the results in our experiment. The experiment data will determine $M_i$, $\tau_i$ and $P_D$. Using this criterion we could confirm whether those data demonstrate steering.

\begin{figure}[h]
\includegraphics[width=0.46\textwidth]{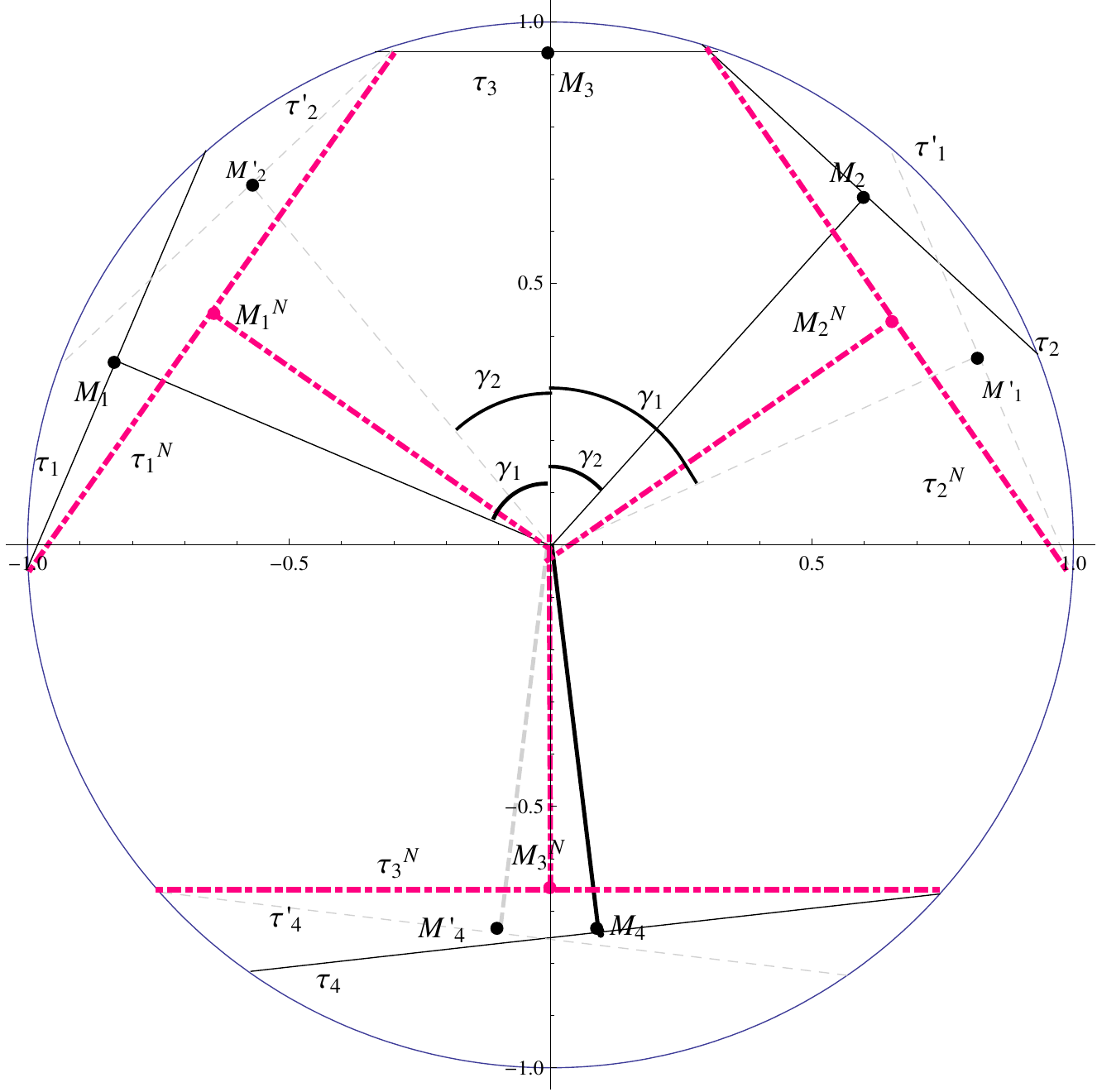}\\
\renewcommand{\thefigure}{S7}
\caption{Symmetrization of the errors. Supposing $\gamma_1 \neq \gamma_2$ and $M_4$ is not on z-axis. Defining $M'_i=\{-M_{ix},M_{iz}\}$ ($i=1,2,4$) and $\tau'_i$ is given by $M'_i$. $\tau^N_1$ is the smallest tangent whose spherical crow contain both $\tau_1$ and $\tau'_2$, $\tau^N_2$ and $\tau^N_4$ are defined similarly.}\label{figure4}
\end{figure}

\subsubsection{Criterion with Experiment Errors.}
Before we using this criterion to our experiment result, we need to deal with the experiment errors first. There are two kinds of errors: the first one is given by the measured values $\langle W_i \rangle$. This type of errors will make $\tau_i$ no more a tangent but has a width which is given by the corresponding error. In the case $r_1 \neq r_2$, notice that all of the lemmas will still hold if we ask $r_{12}=Min\{r_1,r_2\}$. Similarly, the errors of $\langle W_3 \rangle , \langle W_4 \rangle$ only makes $r_3, r_4$ a little bit smaller and the criterion still available. The second kind of errors is given by $\gamma_1 \neq \gamma_2$ and $M_3, M_4$ may not on z-axis. This type of errors will broken the symmetry required in the proof and makes the criterion invalid. While, we could turn the second type of errors to the first type of errors, and make the criterion still available. As Fig.(\ref{figure4}) shows, suppose $\gamma_1 \neq \gamma_2$ and $M_4$ is not on z-axis. (We could always assume $M_3$ is on z-axis, because in all of our proofs we just used the relative locations.) Let us define $M'_i=\{-M_{ix},M_{iz}\}$ ($i=1,2,4$) and $\tau'_i$ is given by $M'_i$. $\tau^N_1$ is the smallest tangent whose spherical crow contain both $\tau_1$ and $\tau'_2$, $\tau^N_2$ and $\tau^N_4$ are defined similarly as show in Fig.(\ref{figure4}), $M^N_i$ is given by $\tau^N_i$. It's easy to see those new $\tau^N_i$ are symmetry as the criterion required. After the symmetrization of the errors we could check whether our experiment data demonstrate the steering nonlocality. The value of $(|OB|-|OG|)_{min}$ minimizes over the region given by the measured value with the corresponding errors. If it larger than zero the state we discussed is steerable.
$\\$

\end{document}